\documentclass[prl,twocolumn,10pt, superscriptaddress]{revtex4-1}
\usepackage{amsfonts}
\usepackage[utf8x]{inputenc}
\usepackage{graphicx}
\usepackage{amsmath} 
\usepackage[caption=false,labelformat=empty,position=top,justification=centering]{subfig}
\begin{document}
\title{Indications of a late-time interaction in the dark sector}
\author{Valentina Salvatelli}
\affiliation{Physics Department and INFN, Universit\`a di Roma ``La Sapienza'', Ple Aldo Moro 2, 00185, Rome, Italy}
\author{Najla Said} 
\affiliation{Physics Department and INFN, Universit\`a di Roma ``La Sapienza'', Ple Aldo Moro 2, 00185, Rome, Italy}
\author{Marco Bruni}
\affiliation{Institute of Cosmology and Gravitation, University of Portsmouth, Dennis Sciama Building, Burnaby Road, Portsmouth PO1 3FX, United Kingdom}
\author{Alessandro Melchiorri}
\affiliation{Physics Department and INFN, Universit\`a di Roma ``La Sapienza'', Ple Aldo Moro 2, 00185, Rome, Italy}
\author{David Wands}
\affiliation{Institute of Cosmology and Gravitation, University of Portsmouth, Dennis Sciama Building, Burnaby Road, Portsmouth PO1 3FX, United Kingdom}

\begin{abstract}
\noindent 
We show that a general late-time interaction between cold dark matter and vacuum energy is favoured by current cosmological datasets. We characterize the strength of the coupling by a dimensionless parameter $q_V$ that is free to take different values in four redshift bins from the primordial epoch up to today.  This interacting scenario is in agreement with measurements of cosmic microwave background temperature anisotropies from the Planck satellite, supernovae Ia from Union 2.1 and redshift space distortions from a number of surveys, as well as with combinations of these different datasets. Our analysis of the 4-bin interaction shows that a non-zero interaction is likely at late times. We then focus on the case $q_V\not=0$ in a single low-redshift bin, obtaining  a nested one parameter extension of the standard $\Lambda$CDM model. We study the Bayesian evidence, with respect to $\Lambda$CDM, of this late-time interaction model, finding moderate evidence for an interaction starting at $z=0.9$, dependent 
upon the prior range chosen for the interaction strength parameter $q_V$. For this case the null interaction ($q_V=0$, i.e.\ $\Lambda$CDM) is excluded at $99\%$ c.l..
 \end{abstract}

\maketitle

\noindent{\bf Introduction:}
Measurements of anisotropies of the cosmic microwave background (CMB) from experiments including the WMAP \cite{Bennett:2012zja} and Planck \cite{Ade:2013zuv} satellites, combined with independent measurements of the cosmic expansion history, such as baryon acoustic oscillations \cite{Anderson:2013zyy}, have provided strong support for the standard model of cosmology with dark energy (specifically a cosmological constant, $\Lambda$) and cold dark matter (CDM). However the latest CMB data are in tension with local measurements of the Hubble expansion rate from supernovae Ia \cite{Riess:2011yx} and other cosmological observables which point towards a lower growth rate of large-scale structure (LSS), including cluster counts \cite{Ade:2013lmv,Vikhlinin:2008ym} and redshift-space distortions (RSD) from galaxy peculiar velocities \cite{Samushia:2012iq}. 

At the present time it remains unclear whether these discrepancies may be due to systematic effects in the different methods used for measurements, or whether they could instead be evidence for deviations from $\Lambda$CDM. 
Massive neutrinos have been proposed to reconcile CMB with LSS observations  \cite{Battye:2013xqa}, but they increase the tension with local measurements of the Hubble rate \cite{Giusarma:2013pmn}.
Dynamical dark energy can help reconcile CMB and local Hubble expansion measurements, but does not ease the tension with LSS  \cite{Ade:2013zuv}.
However, a coupling between the components of the dark sector can strongly influence the evolution of both the background and perturbations. Models with a constant interaction between cold dark matter and dark energy have already been proposed as one possible solution to solve the tension in the measurements of the Hubble constant from CMB and Supernovae \cite{Salvatelli:2013wra}.

In this {Letter} we investigate a minimal extension of the $\Lambda$CDM model where dark matter is allowed to interact with vacuum energy, without introducing any additional degrees of freedom. We allow the interaction strength to vary with redshift and show that energy transfer from dark matter to the vacuum can resolve the tension between the CMB and RSD measurements of the growth of LSS, making it consistent to combine these two datasets. 
We consider only RSD measurements as these probe the gravitational potential in the linear regime and do not depend on non-linear evolution and the formation of collapsed halos and clusters.
Our main result is that a model where an interaction in the dark sector switches on at late times is particularly favoured with respect to $\Lambda$CDM. Assuming an interaction starting at redshift $z=0.9$ the null interaction case (i.e.\ $\Lambda$CDM) is excluded at $99\%$ c.l..
 
\noindent{\bf Model:}
\noindent Interacting vacuum  models (iVCDM) allow energy-momentum transfer between CDM and the vacuum \cite{Bertolami:1986bg,Freese:1986dd,Carvalho:1991ut,Berman:1991zz,Pavon:1991uc,Wands:2012vg}. The background evolution is encoded in the coupled energy conservation equations
\begin{eqnarray}
\dot{\rho}_c + 3H\rho_c&=& -Q, \\
\dot{V}&=&Q ,
\end{eqnarray}
for the CDM and vacuum densities, $\rho_c$ and $V$, the standard conservation equations for baryons, photons and neutrinos, and the  Friedmann equation
\begin{equation}
H^2=\frac{8 \pi G}{3} (\rho_{tot} + V),
\end{equation}
where $\rho_{tot}$ is the total matter and radiation energy density, $H$ is the expansion rate of the universe, $Q$ is the interaction term and we assume a spatially flat universe. When there is no interaction ($Q=0$) we have $8\pi G V=\Lambda$, the cosmological constant, and we recover the standard $\Lambda$CDM model. 

In general the interaction is covariantly represented by a 4-vector $Q^\mu$; if we assume that this is proportional to the 4-velocity of CDM ($Q^{\mu}= Q u^{\mu}$) then the matter flow remains geodesic ($u^\mu\nabla_\mu u^\nu=0$) and in the comoving-synchronous gauge the vacuum energy is spatially homogeneous \cite{Wands:2012vg}. Hence the perturbation equations in this gauge are the same as in $\Lambda$CDM, with zero effective sound speed \cite{Wang:2013qy}.

Recent studies of interacting vacuum cosmologies have focused on specific models for the interaction, $Q(z)$ \cite{DeSantiago:2012xh,Wang:2013qy,Borges:2013bya,Sola:2014tta,Wang:2014xca}. In this {Letter} we want to consider a general interaction $Q(z)$ in different redshift bins. Thus we take an interaction of the form $Q=-q_VHV$, where $q_V(z)$ is a dimensionless parameter that encodes the strength of the coupling  \cite{Quercellini:2008vh}. 
We require $q_V<0$ to ensure that the matter density remains non-negative. Note that in our notation a negative $q_V$ implies dark matter decaying into vacuum. 

We first consider a model in which $q_V(z)$ is a binned (stepwise-defined) function. We have subdivided the redshift range from last scattering until today into four bins, with  $q_V(z)=q_i$ ($i=1..4$), i.e.\ parametrizing our iVCDM model with four parameters. We have chosen to include all the redshifts from the primordial epoch to $z=2.5$ in a single bin (bin 1),  as we have few measurements in that range after CMB last scattering. The other three bins have been chosen with the aim to be mainly sensitive to supernovae (bin 4, $0\leq z\leq 0.3$), to RSD (bin 3, $0.3\leq z\leq 0.9$) and to the farthest supernova observations available (bin 2,  $0.9\leq z\leq 2.5$).

In the light of our results for $q_V(z)$, we then focus on the case of a late-time interaction, with $q_V\not=0$ in a single low-redshift bin.

\noindent{\bf Analysis:} 
\noindent We have performed a Bayesian analysis with the Monte Carlo Markov chain code CosmoMC \cite{Lewis:2002ah,Lewis:2013hha} and a modified version of the Boltzmann code CAMB \cite{Lewis:1999bs}. The datasets we have considered to assess the likelihood of the model are CMB measurements from Planck \cite{Ade:2013kta} including polarization from WMAP \cite{Bennett:2012zja}, SNIa from the compilation Union2.1 \cite{Suzuki:2011hu} and RSD measurements from a number of surveys \cite{Beutler:2012px,Percival:2004fs,Blake:2011rj,Samushia:2011cs,Reid:2012sw,delaTorre:2013rpa}, see Fig.~\ref{fig.bestfit}. We also considered baryon acoustic oscillations \cite{Beutler:2012px,Percival:2009xn,Anderson:2013oza,Blake:2011rj} and radio galaxies data \cite{Daly:2007pp}, finding that the constraints from these datasets are equivalent to those from SN; therefore their addition  to our analysis doesn't change our results. A comprehensive analysis including the  effects of these  datasets will be presented in  a forthcoming paper \cite{Inprep}.
%

In this analysis we have chosen a flat prior [-10,0] for the $q_i$ parameters since the parameters' magnitude is assumed to be of order one. (We will consider later the effect of a wider logarithmic prior, see Fig.~\ref{2D_qvzin} and \cite{Inprep}.)

\begin{figure}[h!]
\includegraphics[trim=1cm 0cm 0cm 0cm, scale=0.35]{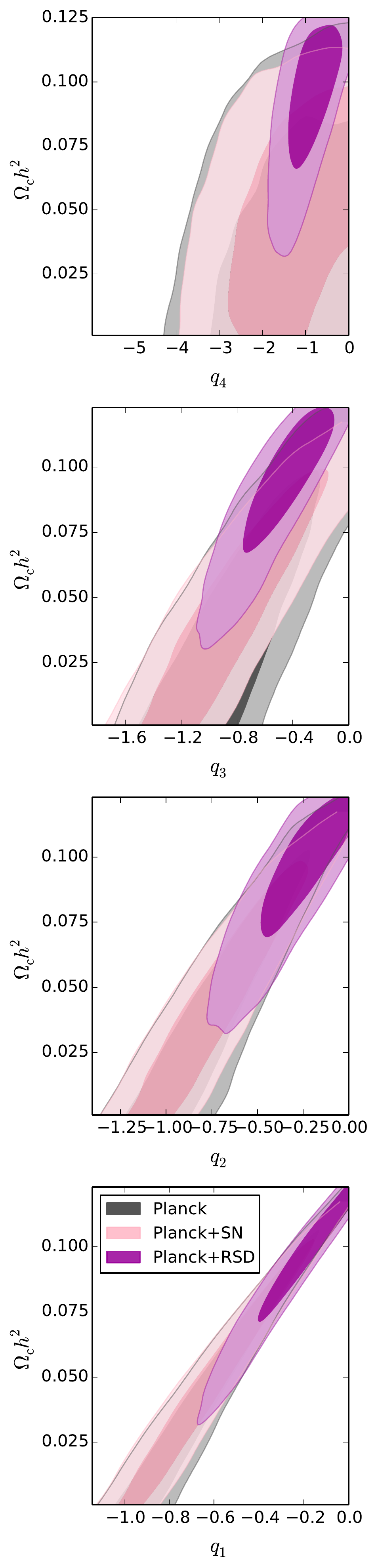}
\includegraphics[trim=0cm 0cm 0cm 0cm, scale=0.35]{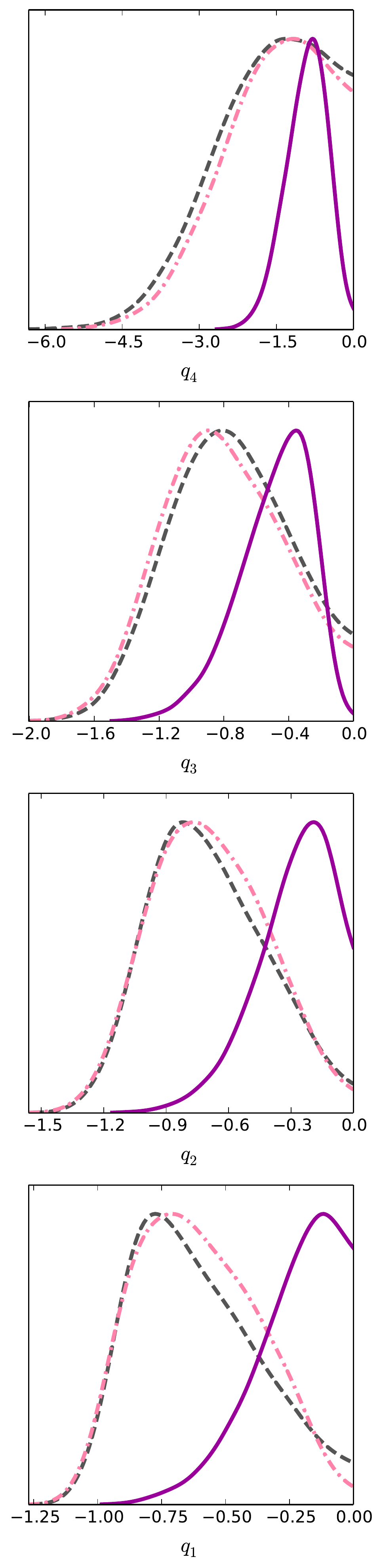}
\caption{Two-dimensional $\Omega_c h^2$-$q_i$ contours at 68 and 95\% c.l. (left) and one-dimensional $q_i$ probability distributions (right) from Planck (black), Planck+SN (pink) and Planck+RSD (purple). The addition of the RSD datasets breaks the degeneracy between the two parameters, $\Omega_c h^2$ and $q_i$, and narrows the probability distributions of $q_3$ and $q_4$ in particular. A null interaction at low-redshift is excluded with high significance.}
\label{fig.4bin}
\end{figure}

\begin{figure}[h!]
\begin{center}
\includegraphics[trim=1.4cm 0.5cm 1.5cm 0.6cm,scale=0.6]{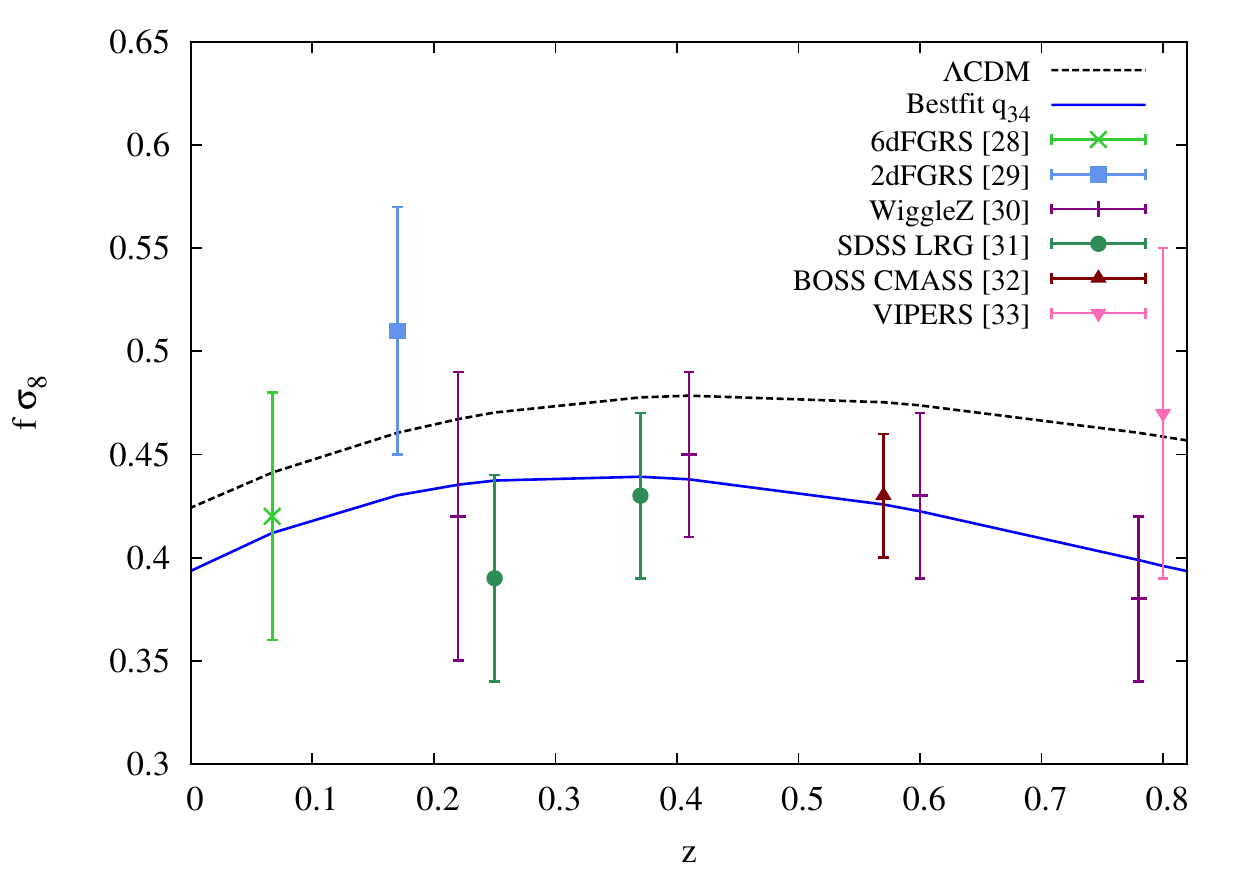}
\caption{RSD measurements \cite{Beutler:2012px,Percival:2004fs,Blake:2011rj,Samushia:2011cs,Reid:2012sw,delaTorre:2013rpa}
plotted against the theoretical predictions from the best-fit iVCDM model with $q_{34}=-0.128$ (blue) and a $\Lambda$CDM model ($q_{34}=0$) with the same values of cosmological parameters (black).}
\label{fig.bestfit}
\end{center}
\end{figure}

In the 4-bin interaction case, when considering CMB only or CMB+SN measurements, the presence of an interaction is allowed but a null interaction is not excluded in any bin (see column 1 and 2 in Table~\ref{tab.4bin}). This is due to the unbroken degeneracy between the strength of the interaction parameters, $q_i$, and the present-day CDM density ($\Omega_{\mathrm{c}} h^2$), shown in Fig.~\ref{fig.4bin}.
\begin{table}
\begin{tabular}{|l||c|c|c|}
 \hline
 & Planck & Planck+SN & Planck+RSD \\
\hline \hline
$100\Omega_b h^2           $& $   2.203\pm0.029$& $   2.203\pm0.029$& $   2.217\pm 0.028$ \\ \hline
$\Omega_c h^2           $& $   < 0.060$& $   0.049_{-   0.044}^{+   0.018}$& $   0.0918_{-   0.010}^{+   0.026}$ \\ \hline
$100\theta_{MC}      $& $   1.0463_{-   0.0024}^{+   0.0032}$& $   1.0460_{-   0.0028}^{+   0.0023}$& $   1.04302_{-   0.00183}^{+   0.00095}$ \\ \hline
$\tau                $& $   0.087_{-   0.014}^{+   0.012}$& $   0.086_{-   0.014}^{+   0.012}$& $   0.086_{-   0.013}^{+   0.012}$ \\ \hline
$n_s                 $& $   0.9597\pm0.0078$& $   0.9599\pm0.0078$& $   0.9638_{-   0.0078}^{+   0.0071}$ \\ \hline 
${\rm{ln}}(10^{10} A_s)  $& $   3.084_{-   0.026}^{+   0.024}$& $   3.082_{-   0.027}^{+   0.024}$& $   3.078\pm0.024$ \\ \hline
$q_1                 $& $  -0.62_{-   0.31}^{+   0.18}$& $  -0.61_{-   0.29}^{+   0.21}$& $ > -0.29$ \\ \hline
$q_2                 $& $  -0.70_{-   0.33}^{+   0.24}$& $  -0.69_{-   0.31}^{+   0.26}$& $  -0.291_{-   0.098}^{+   0.255}$ \\ \hline 
$q_3                 $& $  -0.76_{-   0.40}^{+   0.37}$& $  -0.80_{-   0.42}^{+   0.36}$& $  -0.49_{-   0.16}^{+   0.28}$ \\ \hline
$q_4                 $& $  > -2.12$& $  -1.58_{-   0.506}^{+   1.51}$& $  -0.92_{-   0.34}^{+   0.48}$ \\  
 \hline
\end{tabular}
\caption{Constraints at 68\% c.l. on cosmological parameters in the iVCDM model when $q_V$ is allowed to vary in four redshift bins.}
\label{tab.4bin}
\end{table}

The degeneracy between $\Omega_ch^2$ and the interaction parameters can be broken by the addition of RSD measurements. This imposes a lower limit on the present cold dark matter density and leads to a shift in the posterior distributions for the interaction, as clearly shown in Fig.~\ref{fig.4bin}. A null interaction is then excluded at 99\% c.l. in bin 3 and at 95 \% in bin 4, showing that a late-time interaction is preferred by observations (see column 3 in Table~\ref{tab.4bin}). This result is also supported by a principal component analysis \cite{Inprep}.

We note that the iVCDM model can also alleviate the tension that arises in $\Lambda$CDM between the Hubble constant measurements from Planck ($H_0=67.3 \pm 1.2$) and the Hubble Space Telescope \cite{Riess:2011yx} ($H_0=73.8 \pm 2.4$). The constraint from Planck in the iVCDM case is $H_0=70.4 \pm 2.5$ (see also~\cite{Wang:2014xca}). The combination with RSD measurements breaks the degeneracy between $q_i$ and $\Omega_c h^2$, leading to $H_0=68.0 \pm 2.3$.

In the light of this analysis we have also explored the viability of a simpler model with an interaction that switches on at low redshift whose strength is encoded in a constant $q_V\neq0$ for $z<z_{in}$. In particular, based on the preceding results, we have selected as the interaction starting point $z_{in}=0.9$, i.e., the upper limit of redshift bin 3. For this reason we will refer to it as the $q_{34}$-model.
In this case a null interaction is excluded at 99\% c.l. Results are shown in Table~\ref{tab.q34} and Fig.~\ref{fig.bestfit}.
\begin{table}[htb]
\begin{tabular}{|l||c|c|}
\hline
 & Bestfit & Mean \\
\hline \hline
100$\Omega_{\mathrm{b}} h^2$ & $2.225$ & $2.216\pm0.027$\\ 
\hline
$\Omega_{\mathrm{c}} h^2$ & $0.1170$ & $0.1183\pm0.0023$\\ 
\hline
$100\theta_{\mathrm{MC}}$ & $1.04150$ & $1.04142\pm0.00061$\\ 
\hline
$\tau$ & $0.094$ & $0.087^{+0.012}_{-0.014}$\\ 
\hline
$n_\mathrm{s}$ & $0.9702$ & $0.9633^{+0.0068}_{-0.0067}$\\ 
\hline
$\ln(10^{10} A_\mathrm{s})$ & $3.094$ & $3.080\pm0.024$\\ 
\hline
$q_{34}$ & $-0.128$ & $-0.156^{+0.068}_{-0.056} $\\
\hline
\end{tabular}
\caption{Constraints at $68\%$ c.l. on fundamental cosmological parameters for the iVCDM model with $q_V=q_{34}$.}
\label{tab.q34}
\end{table}

As shown in Table \ref{Tab.chisquared}, when we introduce four parameters to determine the interaction strength in 4 redshift bins, we obtain a much better fit to the data with respect to $\Lambda$CDM. Remarkably, the $q_{34}$-model with a single interaction parameter can match the best fit of the more complex model with 4 independent redshift bins. It reproduces the same best-fit $\chi^2$ with three fewer parameters. 
\begin{table}[htb!]
\begin{center}
\begin{tabular}{|l|c|c|c|c|c|}
\hline
& $\Lambda$CDM & 4 bins & $q_{34}$only  & $q_{34}, z_{in}$ & $m_{\nu}\Lambda$CDM \\ \hline
$\chi^2_{\rm min}$ & 9818 & 9811 & 9811 & 9810 & 9813 \\  \hline
\end{tabular}
\caption{Minimum $\chi^2$ values for CMB+RSD datasets fitted to various theoretical models discussed in the text.}
\label{Tab.chisquared}
\end{center}
\end{table}

An alternative way to compare different models is to compute the Bayesian evidence.
Since the $q_{34}$ model is a one-parameter nested extension of $\Lambda$CDM, we can simply compute the Bayes factor $B_{01}$, that represents the ratio of the models' probability, using the Savage-Dickey Density Ratio formula
\begin{equation}
\label{SDDR}
 B_{01}= 
 \left. \frac{P(q_{34}|data,model1)}{P(q_{34}|model1)} \right|_{q_{34}=0}
\end{equation}
where $q_{34}$ is the additional parameter and $q_{34}=0$ is the value of the parameter for which model 0 ($\Lambda$CDM) is recovered. 

The Bayesian evidence for the extended model is $-\ln B_{01}$;
thus $B_{01}$ less than one means that the $q_{34}$ model is preferred over $\Lambda$CDM.

\begin{figure}[h!]
\begin{center}
\includegraphics[trim=1.5cm 1.0cm 1.5cm 0.8cm,scale=0.32]{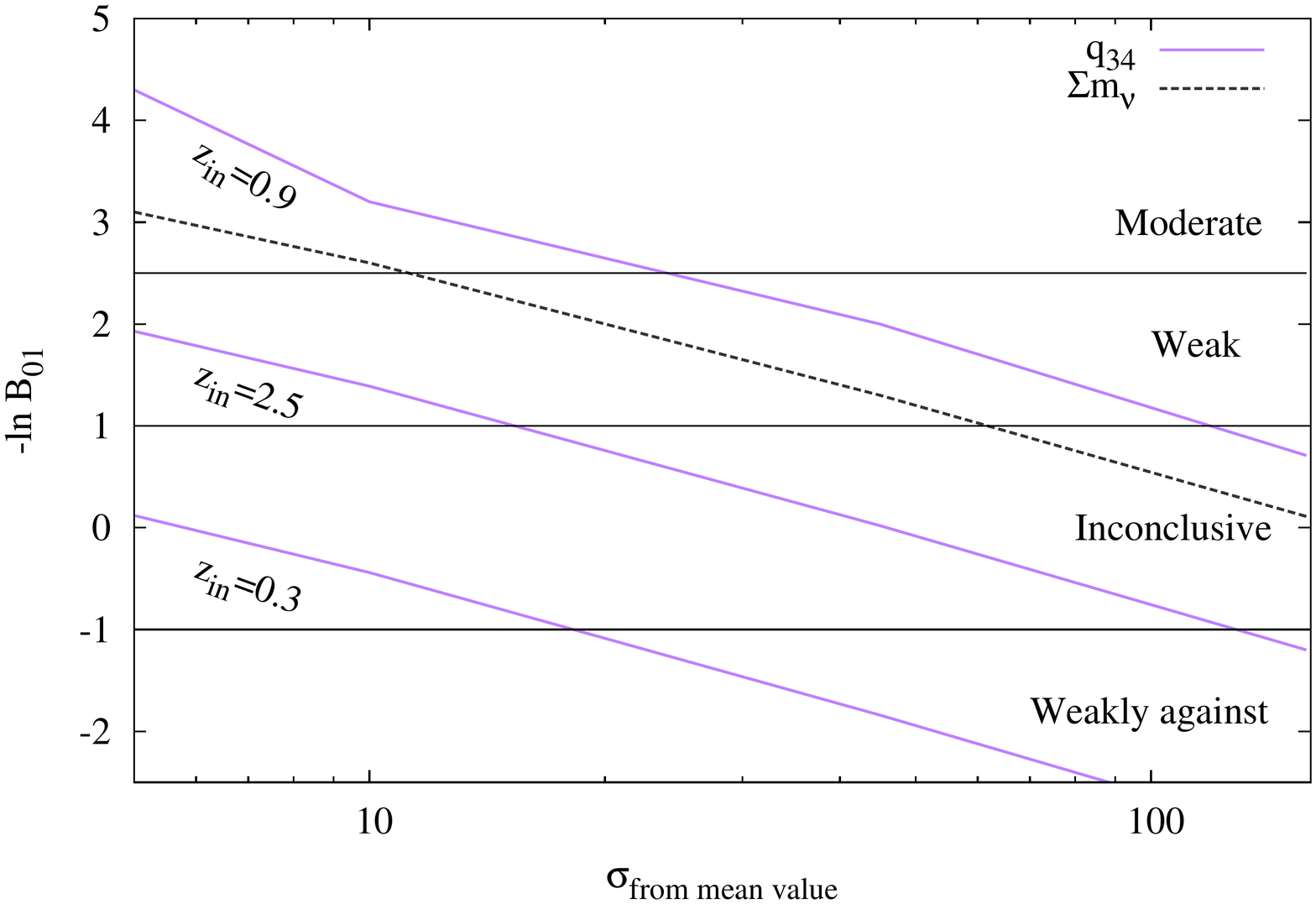} 
\caption{Bayesian evidence as a function of the prior width, expressed in terms of standard deviations from the mean value of the nested parameter. In purple (solid line) the $q_{34}$ model ($z_{in}$=0.9) and same model with different choices of $z_{in}$. In grey (dashed line) the $m_{\nu}$-$\Lambda$CDM model. On the right we report the empirical Jeffreys' scale defined in \cite{Jeffreys:1961}.}
\label{fig.evidence}
\end{center}
\end{figure}

\begin{figure}[h!]
\begin{center}
\subfloat[\newline$\Lambda$CDM \hspace{1cm}]{\includegraphics[trim=3cm 1.5cm 0cm 0.6cm, scale=0.3]{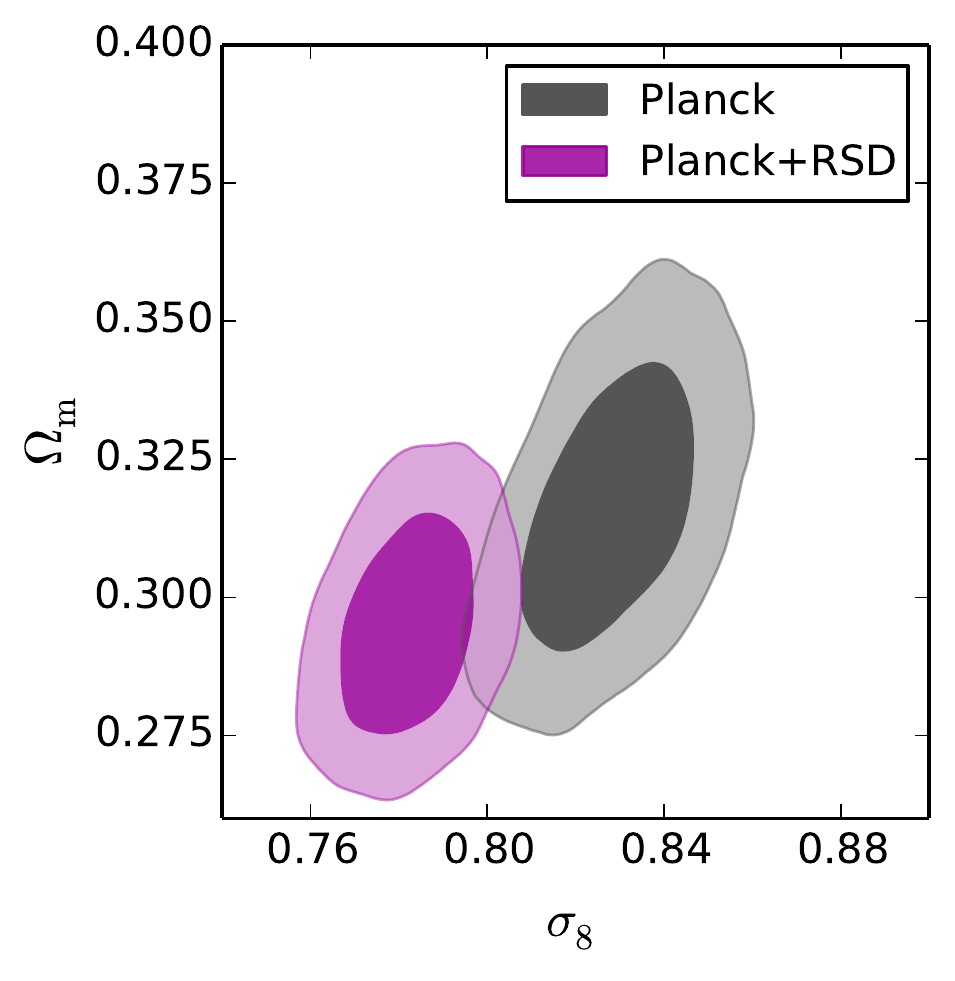}}
\subfloat[\newline iVCDM]{\includegraphics[trim=0cm 1.5cm 0cm 0.6cm,scale=0.3]{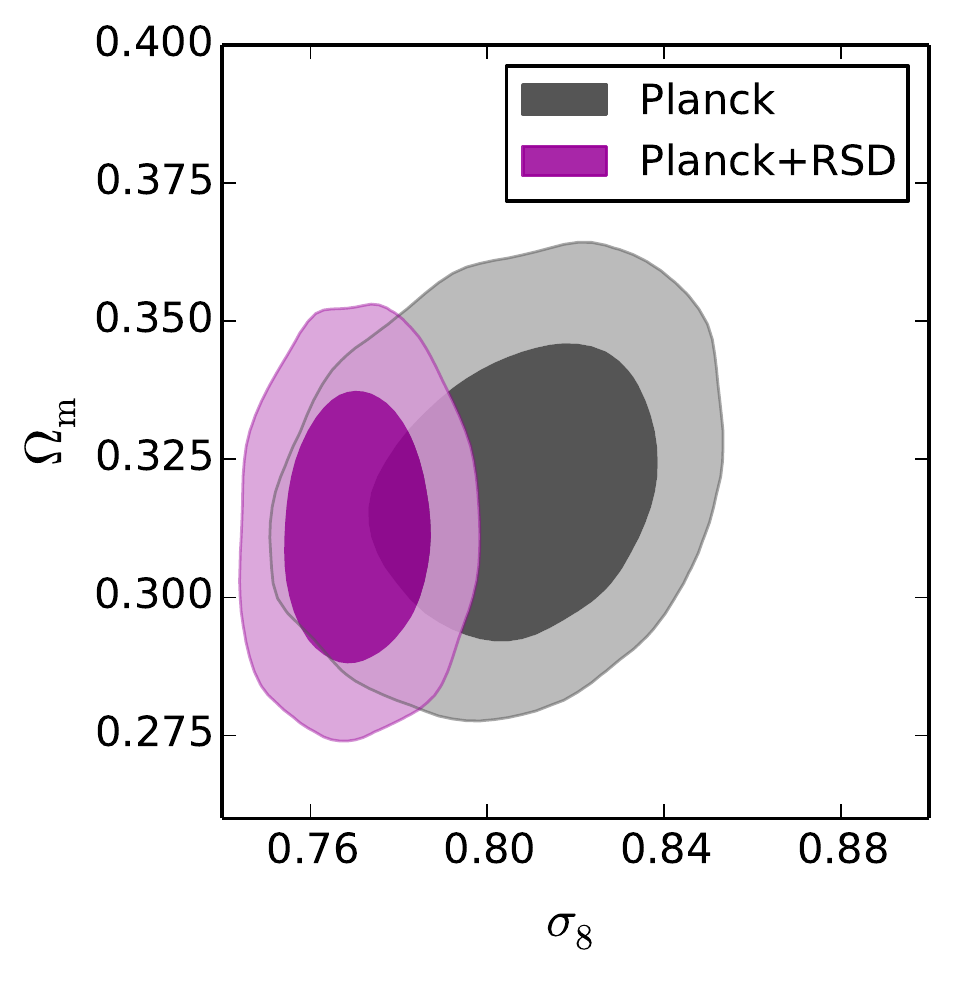}}
\subfloat[\newline $m_{\nu}\Lambda$CDM]{\includegraphics[trim=0cm 1.5cm 2.7cm 0.6cm,scale=0.3]{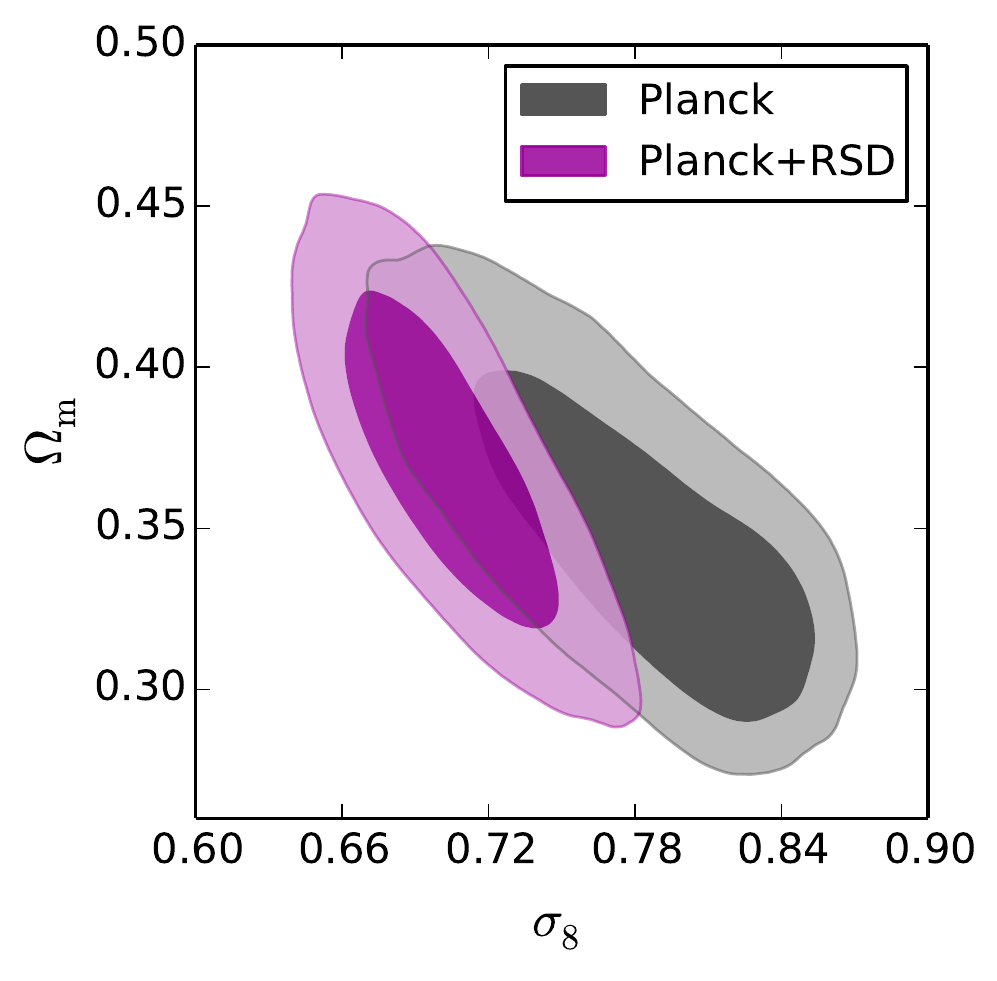}}
\end{center}
\caption{$\Omega_m$-$\sigma_8$ contours at 68 and 95\% c.l. from Planck experiment (black) and Planck+RSD (purple) for three theoretical models. The tension between the Planck and RSD datasets that arises in the $\Lambda$CDM model (left) is resolved in the $q_{34}$ interacting vacuum model (middle). Also in the $\Lambda$CDM model with massive neutrinos (right) this tension with RSD is alleviated (in contrast to the tension that arises when considering non-linear probes of LSS growth \cite{Leistedt:2014sia}).}
\label{fig.2D_tension}
\end{figure}

\begin{figure}[h!]
\begin{center}
\includegraphics[trim=1.5cm 1cm 0cm 1.0cm, scale=0.42]{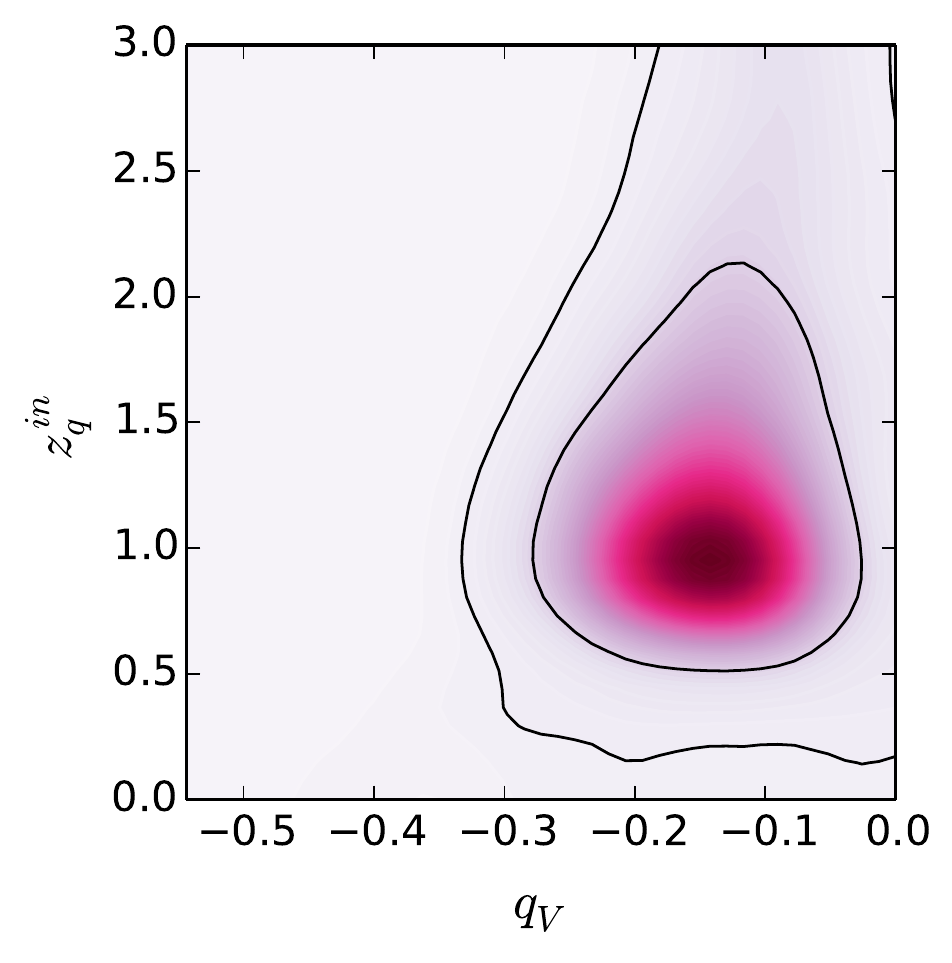}
\includegraphics[trim=4.2cm 3.8cm 5.0cm 3.0cm,scale=0.24]{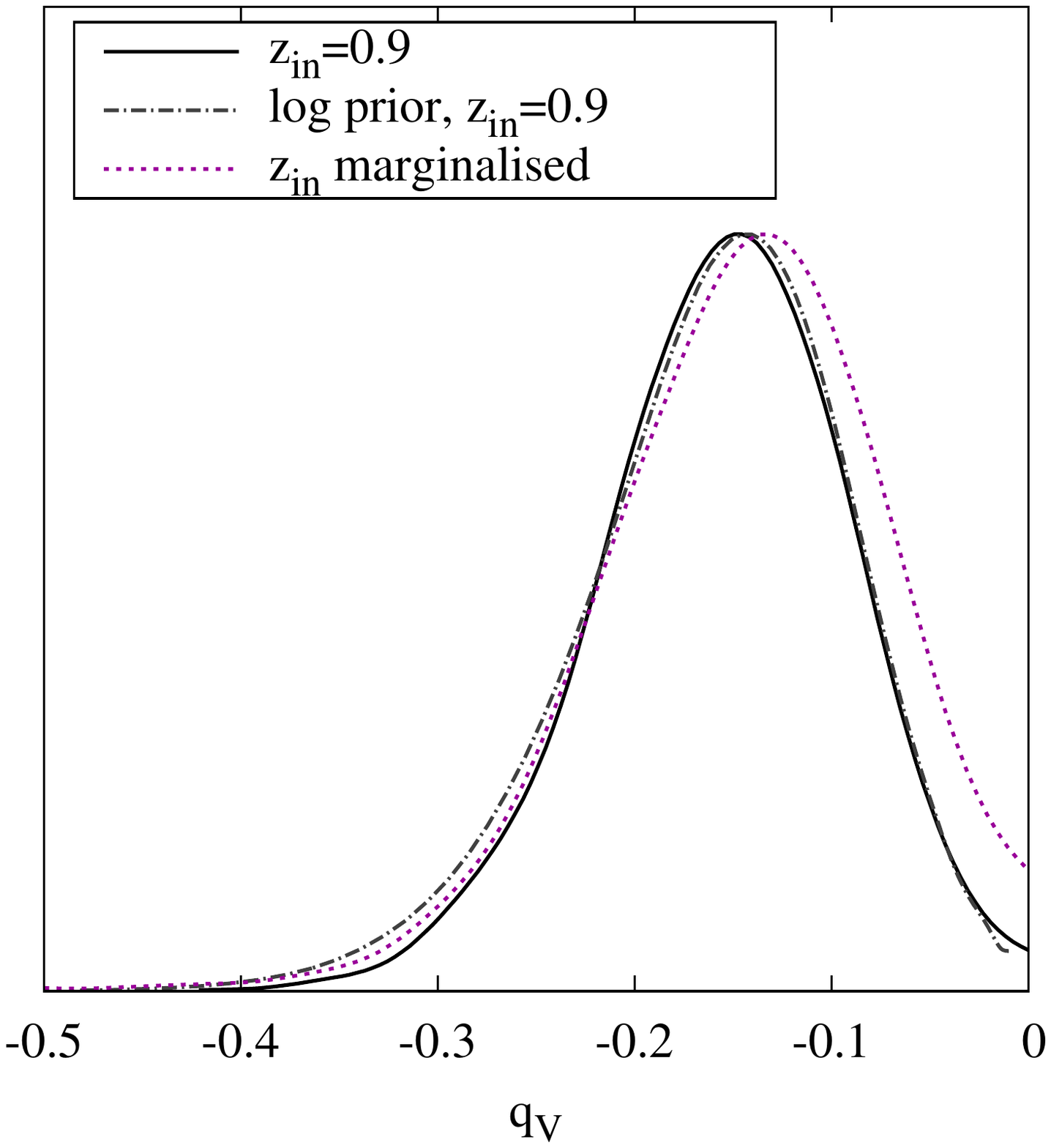}
\caption{Effects of varying  $z_{in}$ in the single bin iVCDM model. Left: $q_V$-$z_{in}$ contours at 68\% and 95\% c.l.;
 $z_{in}$ is poorly constrained but the degeneracy with $q_V$ is weak.
Right: $q_V$ posterior distribution when $z_{in}$ is either fixed at  $z_{in}=0.9$ ($q_{34}$ model) or marginalized. In dashed line the posterior when considering a log prior, $\log_{10}|q_V|\in[-2,2]$ with fixed $z_{in}$.} 
\label{2D_qvzin}
\end{center}
\end{figure}

The Bayesian evidence inevitably depends on the prior distribution of model parameters, decreasing with the prior width. When dealing with phenomenological parameters, such as $q_{34}$, it is not clear what range for the prior should be considered when computing of the evidence \cite{Trotta:2008qt}. For this reason we have explored in Fig.~\ref{fig.evidence} how the evidence changes with the width of the prior. For comparison we have computed the evidence for three late-time interaction models with different choices for $z_{in}$. We have also evaluated the Bayes factor between $\Lambda$CDM and another one parameter extension that alleviates the tension between CMB and RSD measurements, namely $\Lambda$CDM with massive neutrinos \cite{Battye:2013xqa}, see Fig.~\ref{fig.2D_tension}. In this case the nested extra parameter is the sum of the neutrino masses, for which we obtain $\sum m_{\nu}=0.53\pm 0.19$ eV. In the rest of our analysis we use the standard fixed value $\sum m_{\nu}=0.06$ eV. We see that the evidence for 
the $q_{34}$ 
model with $z_{in}=0.9$ is always higher than the other one-parameter models we study for a given prior width relative to the standard deviation from the mean. The evidence remains moderate even when allowing a prior range for $q_{34}$ equal to $20$ standard deviations from the mean.

\noindent{\bf Possible biases:}
In order to check the robustness of our results we have performed some further analysis \cite{Inprep}. 
First we have explored a model where $z_{in}$ is free to vary. Table \ref{Tab.chisquared} 
shows that $z_{in}=0.9$ is a good approximation of the best-fit point of this extended model, and hence maximises the Bayesian evidence computed above. 
Marginalizing over $z_{in}$ slightly broadens and shifts the posterior distribution for $q_V$, as shown in Fig.~\ref{2D_qvzin}.
We also show in Fig.~\ref{2D_qvzin} that choosing a wider logarithmic prior, $\log_{10}|q_{34}|\in[-2,+2]$ for fixed $z_{in}$, has a small effect on the posterior.

Moreover we have tested our results against variation of the lensing amplitude parameter of the CMB temperature, $A_L$. In $\Lambda$CDM Planck measurements point towards an $A_L$ value that is higher than the standard value, $A_L=1$, \cite{Ade:2013zuv,Said:2013hta} used in the preceding analysis. 
In our iVCDM model a degeneracy exists between $A_L$ and $q_V$ that reduces the strength of the interaction when $A_L$ increases. However the indication for an interaction is maintained at 95\% c.l..

Finally, given the recent results from the BICEP2 experiment that claims a detection for a tensor to scalar ratio $r$ different from zero \cite{Ade:2014xna}, we have investigated if our results may be affected. The interaction parameter is actually very poorly degenerate with $r$
and the inclusion of the BICEP2 dataset changes the Bayes evidence by only 1\%. 

\noindent{\bf Conclusions:}
We have shown that an interacting vacuum cosmology, where the strength of the coupling  with CDM varies with redshift, is a possible solution to the tension that arises in the standard $\Lambda$CDM model between CMB data and LSS linear growth measured by RSD, see Fig.~\ref{fig.2D_tension}. In particular we have found that an interaction which switches on at late times ($z\sim0.9$) is particularly favoured. In this context we have obtained a very tight constraint on the interaction strength parameter, excluding the $\Lambda$CDM model (i.e., a null interaction) at 99\% c.l.. We have also verified that the probability of late-time interaction is only weakly affected by changes in the value of the tensor-to-scalar ratio or the lensing amplitude parameters.

We have only considered here constraints on the linear growth of LSS, as the non-linear coupled evolution of interacting vacuum and dark matter has yet to be studied in detail. It will be important to examine the predictions of iVCDM for cluster number counts; this provides tight constraints, e.g.\ on $\Lambda$CDM, but requires non-linear modelling. Interacting vacuum models can be recast \cite{Wands:2012vg}  as  clustering quintessence with vanishing sound speed \cite{Creminelli:2009mu} and/or irrotational dark matter \cite{Sawicki:2013wja}, either of which could have distinctive predictions for non-linear collapse.
\\
\begin{acknowledgments}
\noindent{\bf Acknowledgments}
We would like to thank Rob Crittenden, Karen Masters, Lado Samushia and Gong-Bo Zhao for helpful comments. The work of MB and DW was supported by STFC grants ST/K00090X/1 and ST/L005573/1. VS and NS are grateful to the ICG for its hospitality. We thank an anonymous 
referee for suggesting the use of radio galaxies data \cite{Daly:2007pp}.
\end{acknowledgments}

\end{document}